\documentstyle[12pt]{article}
\newcommand{\fig}[1]{Fig.\ref{#1}}
\renewcommand{\a}[3]{ a_{ {#2}{#3} }({#1})}
\renewcommand{\b}[2]{ b_{ {#2} }({#1})}
\renewcommand{\c}[2]{ c_{ {#2} }({#1})}
\newcommand{\nn}{\nonumber}
\newcommand{\ns}{\normalsize}
\newcommand{\bc}{\begin{center}}
\newcommand{\ec}{\end{center}}
\newcommand{\lb}{\linebreak}
\newcommand{\bq}{\begin{equation}}
\newcommand{\eq}{\end{equation}}
\newcommand{\bqa}{\begin{eqnarray}}
\newcommand{\eqa}{\end{eqnarray}}
\newcommand{\ben}{\begin{enumerate}}
\newcommand{\een}{\end{enumerate}}
\newcommand{\eqn}[1]{Eq.(\ref{#1})}
\def\notp{p\hspace*{-4.8pt}/}
\def\notq{q\hspace*{-5.8pt}/}

\def\l{\lambda}
\def\F{{\cal F}}
\def\ep{\epsilon}

\def\D{{\cal D}}
\def\G{{\cal G}}
\def\bw{\Omega}
\def\w{\omega}\def\wa{\omega_1}\def\wb{\omega_2}
\def\wc{\omega_3}\def\wab{\omega_{12}}
\newcommand{\ps}[1]{\psi_{\mathrm{#1}}}
\begin{document}
\input feynman
\pagestyle{empty}

\begin{flushright}CERN-TH.6853/93\\
\end{flushright}
\vspace*{1cm}
\begin{center}\begin{Large}
{\bf Nullification of multi-Higgs threshold
amplitudes in the Standard Model}\end{Large}

\vspace*{2cm}

 E.N.~Argyres   \\
 {\ns Institute of Nuclear Physics, NRCPS
 `$\Delta  \eta  \mu  \acute{o}  \kappa  \varrho
   \iota  \tau  o  \varsigma$', Greece}\\
 \vspace{\baselineskip}
 Ronald~H.P.~Kleiss,\\
 {\ns NIKEHF-H, Amsterdam, the Netherlands}\\
 \vspace{\baselineskip}
Costas~G.~Papadopoulos\\
{\ns TH Division, CERN, Geneva, Switzerland}\\
\vspace*{3cm}
Abstract\\[24pt] \ec
We show that nullification of all tree-order threshold
amplitudes involving Higgs particles in the Standard Model occurs,
provided that certain equations relating the masses
of all existing elementary particles to the mass
of the Higgs scalar are satisfied.
The possible role of these relations
in restoring the high-multiplicity unitarity and their
phenomenological relevance are briefly discussed.
\vspace{1cm}
\begin{flushleft} CERN-TH.6853/93\\March 1993\\
\vfill
\end{flushleft}

\newpage
\pagestyle{plain}
\setcounter{page}{1}

\par
The {\it high-multiplicity} limit of processes involving scalar
particles has been studied recently \cite{volo:1,akpa:1}. At tree order
the amplitudes ${\cal A}(H^*\to nH)$ as well as the
cross section $\sigma(f\bar{f}\to nH)$, to leading order in the
Yukawa coupling \cite{akpa:4,volo:2},
grow as $n!$, where $n$ is the number of
produced scalar particles, and violate the unitarity bounds
for sufficiently high energies. The reason for this behaviour is
rather simple and relies on the coherence properties of
scalar amplitudes: all amplitudes at tree order add coherently,
so the $n!$ simply counts the number of Feynman diagrams.
\par The study of the
high-multiplicity limit of amplitudes, within the framework
of perturbative quantum field theory, might have a
profound theoretical and phenomenological
interest. The situation resembles that of
the {\it high-energy} limit of amplitudes involving longitudinal
bosons. As is well known \cite{tikt:1}, the latter amplitudes
violate the unitarity bound, unless specific relations hold among
the different couplings. The existence
of such relations causes very delicate cancellations at tree order,
which are responsible for the restoration of high-energy unitarity.
The persistence of these cancellations at higher orders is
naturally understood
in the framework of the $SU(2)_L\times U(1)_Y$ gauge symmetry,
which guarantees a consistent behaviour of the
amplitudes in the high-energy limit, by discarding from all physical
processes the `unphysical' Goldstone bosons.
Nevertheless, since in the high-energy limit ($\sqrt{s}\to \infty,
n\;\mathrm{fixed}$)
one can neglect any mass dependence, no direct predictions can
be made concerning the masses\footnote{Except for the ratio $m_W/m_Z$.}.
The situation is opposite in the case of the high-multiplicity limit
($\sqrt{s}\to \infty, \sqrt{s}/nm\;\mathrm{fixed}$),
since the unitarity violation is now related to the threshold
behaviour of the amplitudes, so that the mass dependence
can no longer be neglected.
As we will show, it is the advantage of the
high-$n$ limit of multi-Higgs amplitudes at threshold
that confronts the mass
parameters with the unitarity limits, and that it
provides, in principle,
a way to directly extract information on the masses in the framework
of the Standard Model.
\par In this letter we investigate the possibility that,
given certain relations among the masses of fermions, Higgs particle
and gauge bosons, all physical multiboson amplitudes respect unitarity
at tree order in the high-$n$ limit.
The idea is rather simple:
we seek a mechanism that may discard the bad high-$n$
behaviour of the amplitude ${\cal A}(H^*\to nH)$
from all physical processes.
\par First of all it is easy to see that the physical
amplitudes ${\cal A}(HH\to nH)$ \lb \cite{volo:3,akpa:2,akpa:3}
not only do not exhibit
factorial growth, but  are actually zero (!) for
$n\ge 3$. As was shown in ref.\cite{akpa:3} this nullification
phenomenon is independent of the values of the mass and the self-coupling
of the scalar particle, and depends only on the specific form
of the interaction. It is worth while to recall that the only
interaction exhibiting the nullification phenomenon and
incorporating $\phi^3$ as well as $\phi^4$ terms, is the
spontaneously broken $\phi^4$ theory.
\par The next step is to consider the couplings with the
fermions. The recursion relation for the amplitudes
$d(n)\equiv{\cal A}(f(p)\bar{f}(p')\to nH)$
is given by (see \fig{fi02}):
\begin{figure}[th]
\begin{picture}(16000,20000)

\drawline\fermion[\NE\REG](12000,5000)[6000]
\global\advance\pbackx by 1414
\global\advance\pbacky by 1414
\put(\pbackx,\pbacky){\circle{4000}}
\global\advance\pbacky by -300
\global\advance\pbackx by -450
\put(\pbackx,\pbacky){$n$}
\global\advance\pbackx by 450
\global\advance\pbacky by 300

\global\advance\pbackx by -1414
\global\advance\pbacky by 1414
\drawline\fermion[\NW\REG](\pbackx,\pbacky)[6000]
\global\advance\pfrontx by 1414
\global\advance\pfronty by -1414

\global\advance\pfrontx by 4000
\global\advance\pfronty by -300
\global\advance\pfrontx by -450
\put(\pfrontx,\pfronty){$=$}
\global\advance\pfrontx by 450
\global\advance\pfronty by 300

\global\advance\pfrontx by 7000
\global\newcount\px
\global\newcount\py
\px=\pfrontx
\py=\pfronty

\drawline\fermion[\N\REG](\pfrontx,\pfronty)[3000]
\global\advance\pbacky by 2000
\put(\pbackx,\pbacky){\circle{4000}}
\global\advance\pbacky by -300
\global\advance\pbackx by -450
\put(\pbackx,\pbacky){$n_1$}
\global\advance\pbackx by 450
\global\advance\pbacky by 300

\global\advance\pbacky by 1414
\global\advance\pbackx by -1414
\drawline\fermion[\NW\REG](\pbackx,\pbacky)[4000]

\drawline\fermion[\S\REG](\px,\py)[3000]
\drawline\fermion[\SW\REG](\pbackx,\pbacky)[6000]

\drawline\scalar[\SE\REG](\pfrontx,\pfronty)[2]
\global\advance\pbackx by 1414
\global\advance\pbacky by -1414
\put(\pbackx,\pbacky){\circle{4000}}
\global\advance\pbacky by -300
\global\advance\pbackx by -450
\put(\pbackx,\pbacky){$n_2$}
\global\advance\pbackx by 450
\global\advance\pbacky by 300

\end{picture}
\caption[.]{Diagrammatic representation of the recursion relation
for the process \lb $f\bar{f}\to n H$.
The blob connected with a dashed line corresponds to
the amplitude $H^*\to nH$.}
\label{fi02}
\end{figure}
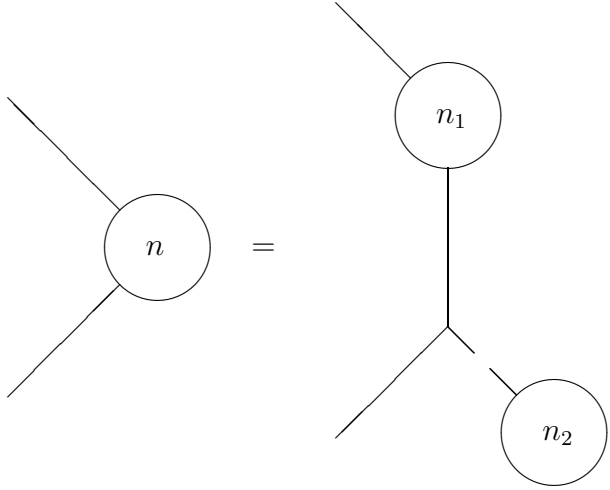
\bq {d(n)\over n!} = -ig_Y\sum\limits_{\begin{small}
\begin{array}{c}
n_{1}\ge 0,n_{2}\ge 1 \\  n_{1}+n_{2}=n
\end{array}
\end{small} }
{id(n_1)\over n_1!P(n_1)}
{ia(n_2)\over n_2!(n_2^2-1)}
\eq
where $P(n)=n\notq-\notp-m$, $m$ is the mass of the fermion and
$g_Y$ the Yukawa coupling.
Throughout this paper
we assume $m_H=1$, and we restore the $m_H$ dependence only when
this is necessary.
Defining
\bq d(n)=-in!\tilde{d}(n) P(n)\;
\;\mathrm{and}\;f=\sum_{n\ge 0}\tilde{d}(n)x^n\eq
and using that \cite{akpa:2}
\bq a(n)=-in!(n^2-1)b(n)\;,
\;f_0(x)=\sum b(n)x^n={x\over 1-\sqrt{\l\over 12}x}
\eq
we have
\bq xf'(x)\notq-f(x)(\notp+m)-2m{z\over 1-z}f(x)=0\eq
where $z=\sqrt{\l/12}\;x$, $d(1)=-ig_Y\bar{u}(p)$ and $f(0)=\bar{u}(p)$.
Taking now $y=-z/(1-z)$ and  $(f(x)=h(y))$, we find
\bq y(1-y)h'(y)\notq-h(y)(\notp+m)+2 m y h(y)=0\eq
Writing $h(y)=\alpha(y)\bar{u}(p)+\beta(y)\bar{u}(p)\notq$
we obtain
\bqa y(1-y)\beta'(y)-2p\cdot q\beta (y)+2my \alpha(y)&=&0\nn\\
-2m\beta(y)+y\alpha'(y)&=&0 \eqa
which gives rise to
\bqa y(1-y)\alpha''(y)+\alpha'(y)(c-y)+4m^2\alpha(y)=0
\eqa
with $c=-2p\cdot q+1$, which is similar to the equations obtained
for the amplitudes \lb ${\cal A}(HH\to nH)$ \cite{akpa:2,akpa:3}.
The results are
\bqa \alpha(y)&=& F(\nu,-\nu;c;y)\nn\\
\beta(y)&=&-{2m\over c}yF(1+\nu,1-\nu;1+c;y)\eqa
with $\nu=2m$ and $F(a,b;c;z)$ is the hypergeometric function.
It is easy to see that the amplitude
${\cal A}(f\bar{f}\to nH)$ vanishes for any $n\ge N$ provided that
\bq {2m_f\over m_H}=N\;\;\;\label{masf}\eq
where $N$ is a non-negative integer \cite{volo:4}.
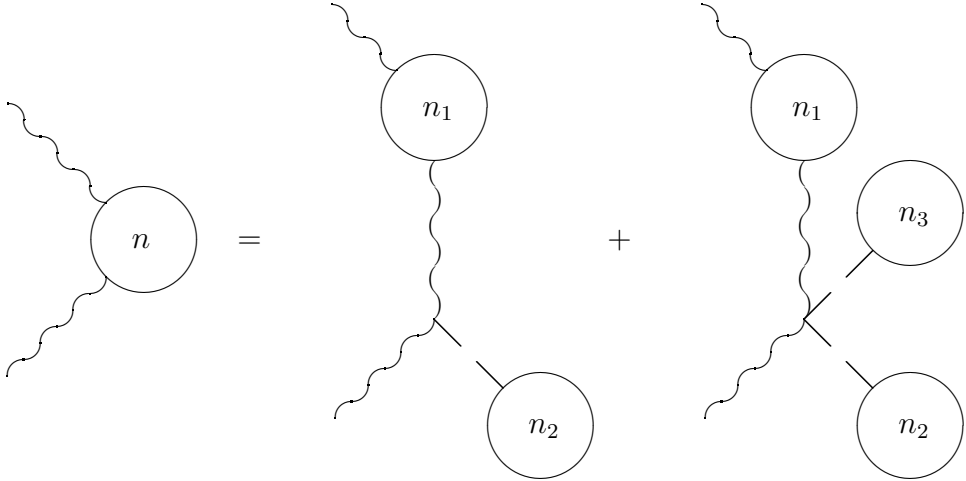
\begin{figure}[th]
\begin{picture}(20000,20000)

\drawline\photon[\NE\REG](3000,6000)[6]
\global\advance\pbackx by 1414
\global\advance\pbacky by 1414
\put(\pbackx,\pbacky){\circle{4000}}
\global\advance\pbacky by -300
\global\advance\pbackx by -450
\put(\pbackx,\pbacky){$n$}
\global\advance\pbackx by 450
\global\advance\pbacky by 300

\global\advance\pbackx by -1414
\global\advance\pbacky by 1414
\drawline\photon[\NW\REG](\pbackx,\pbacky)[6]
\global\advance\pfrontx by 1414
\global\advance\pfronty by -1414

\global\advance\pfrontx by 4000
\global\advance\pfronty by -300
\global\advance\pfrontx by -450
\put(\pfrontx,\pfronty){$=$}
\global\advance\pfrontx by 450
\global\advance\pfronty by 300

\global\advance\pfrontx by 7000
\global\newcount\px
\global\newcount\py
\px=\pfrontx
\py=\pfronty

\drawline\photon[\N\REG](\pfrontx,\pfronty)[3]
\global\advance\pbacky by 2000
\put(\pbackx,\pbacky){\circle{4000}}
\global\advance\pbacky by -300
\global\advance\pbackx by -450
\put(\pbackx,\pbacky){$n_1$}
\global\advance\pbackx by 450
\global\advance\pbacky by 300

\global\advance\pbacky by 1414
\global\advance\pbackx by -1414
\drawline\photon[\NW\REG](\pbackx,\pbacky)[4]

\drawline\photon[\S\REG](\px,\py)[3]
\drawline\photon[\SW\REG](\pbackx,\pbacky)[6]

\drawline\scalar[\SE\REG](\pfrontx,\pfronty)[2]
\global\advance\pbackx by 1414
\global\advance\pbacky by -1414
\put(\pbackx,\pbacky){\circle{4000}}
\global\advance\pbacky by -300
\global\advance\pbackx by -450
\put(\pbackx,\pbacky){$n_2$}
\global\advance\pbackx by 450
\global\advance\pbacky by 300

\global\advance\px by 7000

\global\advance\py by -300
\global\advance\px by -450
\put(\px,\py){$+$}
\global\advance\px by 450
\global\advance\py by 300

\global\advance\px by 7000
\pfrontx=\px
\pfronty=\py

\drawline\photon[\N\REG](\pfrontx,\pfronty)[3]
\global\advance\pbacky by 2000
\put(\pbackx,\pbacky){\circle{4000}}
\global\advance\pbacky by -300
\global\advance\pbackx by -450
\put(\pbackx,\pbacky){$n_1$}
\global\advance\pbackx by 450
\global\advance\pbacky by 300

\global\advance\pbacky by 1414
\global\advance\pbackx by -1414
\drawline\photon[\NW\REG](\pbackx,\pbacky)[4]

\drawline\photon[\S\REG](\px,\py)[3]
\drawline\photon[\SW\REG](\pbackx,\pbacky)[6]

\drawline\scalar[\SE\REG](\pfrontx,\pfronty)[2]
\global\advance\pbackx by 1414
\global\advance\pbacky by -1414
\put(\pbackx,\pbacky){\circle{4000}}
\global\advance\pbacky by -300
\global\advance\pbackx by -450
\put(\pbackx,\pbacky){$n_2$}
\global\advance\pbackx by 450
\global\advance\pbacky by 300

\drawline\scalar[\NE\REG](\pfrontx,\pfronty)[2]
\global\advance\pbackx by 1414
\global\advance\pbacky by 1414
\put(\pbackx,\pbacky){\circle{4000}}
\global\advance\pbacky by -300
\global\advance\pbackx by -450
\put(\pbackx,\pbacky){$n_3$}
\global\advance\pbackx by 450
\global\advance\pbacky by 300

\end{picture}

\caption[.]{Diagrammatic representation of
the recursion formula for the amplitude \lb $VV\to nH$ at threshold.
The blobs connected with a dashed line corresponds to
the amplitude $H^*\to nH$.}
\label{fi01}
\end{figure}
\par We now turn to the amplitudes $VV\to nH$, where $V$
stands for either a $W^\pm$ or a $Z^0$.
The recursion relation for
\bq{\cal A}(V(k_1)V(k_2)\to nH(q) )=
\a{n}{\mu}{\nu} \ep^\mu(k_1;\l_1)\ep^\nu(k_2;\l_2)\eq
is given by (see \fig{fi01}):
\bqa
{\a{n}{\mu}{\nu}\over n!} &=& igM
\sum\limits_{ {\footnotesize \begin{array}{c}
n_{1}\ge 0, n_{2}\ge 1 \\  n_{1}+n_{2}=n \end{array} }}
{\cal P}_\nu^\rho (n_1){ i \a{n_1}{\rho}{\mu}\over n_1!}
{i a(n_2)\over n_2!(n_2^2-1)}\nn\\
&+& i{g^2\over 4}
\sum\limits_{{\footnotesize \begin{array}{c}
n_{1}\ge 0, n_2,n_3\ge 1 \\  n_1+n_2+n_3=n \end{array} }}
{\cal P}_\nu^\rho (n_1){ i \a{n_1}{\rho}{\mu}\over n_1!}
{i a(n_2)\over n_2!(n_2^2-1)}
{i a(n_3)\over n_3!(n_3^2-1)}\label{recr}\eqa
where
\bq {\cal P}_{\mu\nu}(n)= {1\over Q^2-M^2}
\biggl( -g_{\mu\nu}+{ Q_\nu Q_\mu \over M^2 }\biggr)   \eq
$Q^\mu= k_1^\mu-nq^\mu$, $k_1=(E;\vec{k_1})$,
$g$ is the gauge coupling
constant and $M=m_W$.
\par Since the only momenta available are $k_1$ and
$q$, the general form of $\a{n}{\mu}{\nu}$, taking
into account that $k_1\cdot \ep(k_1;\l_1)=0$, is given by:
\bq \a{n}{\mu}{\nu}=a_1(n) g_{\mu\nu}+a_2(n) k_{1\nu} q_\mu +
a_3(n) q_\nu q_\mu\;. \eq
For transverse $V$'s it is easy to see that only $a_1$ survives
after the contraction with the polarization vectors. The
equation for $a_1$ is:
\bqa
{a_1(n)\over n!} &=&-igM
\sum\limits_{ {\footnotesize \begin{array}{c}
n_{1}\ge 0, n_{2}\ge 1 \\  n_{1}+n_{2}=n \end{array} }}
{i a_1(n_1)\over n_1!(n_1^2-1)}
{i a(n_2)\over n_2!(n_2^2-1)}\nn\\
&-& i{g^2\over 4}
\sum\limits_{{\footnotesize \begin{array}{c}
n_{1}\ge 0, n_2,n_3\ge 1 \\  n_1+n_2+n_3=n \end{array} }}
{i a_1(n_1)\over n_1!(n_1^2-1)}
{i a(n_2)\over n_2!(n_2^2-1)}
{i a(n_3)\over n_3!(n_3^2-1)}\label{rec1}\eqa
The above equation has been studied extensively
in refs. \cite{akpa:2,akpa:3}
and it leads to the nullification of $a_1(n)$ for $n>N$ provided that
\bq {4m_W^2\over m_H^2}=N(N+1)\;\;\; .\label{masw}\eq
\par For longitudinal $V$'s, things are slightly more
complicated. Defining

\bq \b{n}{\mu}=\a{n}{\mu}{\nu}\ep^\nu(k_1;\l_1=0)\eq
we have
\bqa
{\b{n}{\mu}\over n!} &=& igM
\sum\limits_{ {\footnotesize \begin{array}{c}
n_{1}\ge 0, n_{2}\ge 1 \\  n_{1}+n_{2}=n \end{array} }}
{\cal P}_\nu^\rho (n_1){ i \b{n_1}{\rho}\over n_1!}
{i a(n_2)\over n_2!(n_2^2-1)}\nn\\
&& + i{g^2\over 4}
\sum\limits_{{\footnotesize \begin{array}{c}
n_{1}\ge 0, n_2,n_3\ge 1 \\  n_1+n_2+n_3=n \end{array} }}
{\cal P}_\nu^\rho (n_1){ i \b{n_1}{\rho}\over n_1!}
{i a(n_2)\over n_2!(n_2^2-1)}
{i a(n_3)\over n_3!(n_3^2-1)}\;\;.\label{rec2}\eqa
Using the obvious ansatz:
\bq \b{n}{\mu}= i n! \biggl( (Q^2-M^2)g_\mu^\nu-Q_\mu Q^\nu \biggr)
\c{n}{\nu} \eq
and writing $\c{n}{\mu}=c_1(n) k_{1\mu} +c_2(n) q_\mu $ we obtain
the following system of recursion relations:
\bqa
(n^2-En-M^2)c_1(n)+(n-E)c_2(n) &=& 4M^2
\sum_{k=0}^{n-1} c_1(k) (n-k)\biggl({\l\over 12}\biggr)^{(n-k)/2}
\nn\\
n(M^2-nE)c_1(n)-nEc_2(n) &=& 4M^2
\sum_{k=0}^{n-1} c_2(k) (n-k)\biggl({\l\over 12}\biggr)^{(n-k)/2}
\nn\eqa
Defining as usual
\bq  f_1(x)=\sum_{n\ge 0} c_1(n) x^n\;\;\mathrm{and}\;\;
f_2(x)=\sum_{n\ge 0} c_2(n) x^n \eq
we arrive at the following system of second-order
differential equations:
\bqa
(\D^2-E\D-M^2)f_1+(\D-E)f_2 &=& {4M^2z\over(1-z)^2}f_1\\
(-E\D^2+M^2\D)f_1-E\D f_2 &=& {4M^2z\over(1-z)^2}f_2\eqa
where $\D$ is the differential operator:
\bq \D\equiv x{d\over dx} \;\;.\eq
Defining $x=-\sqrt{12/\l}\;e^{2\tau}$ and
\bq G=e^{-2E\tau}\biggl({1-z\over 1+z}\biggr)(\D f_1+f_2)\label{gdef}
\eq
we arrive after some straightforward manipulations, at the equation
\bq
\biggl({d^2\over d\tau^2}
 -4E^2 +{4 M^2\over \cosh^2\tau}-{2\over \sinh^2\tau}\biggr)
G=0\;\;.\label{posc}\eq
This is nothing but the the Schr\"odinger equation with
a P\"oschl-Teller potential.
The explicit (but ugly) form of the $G$ is given in ref.\cite{baru:1},
from which the explicit form of $f_{1,2}$ can be derived (see also
\eqn{gdef}), taking into account that
\bq f_1= {1\over M^2} \biggl({1-z\over 1+z}\biggr)^2
\D \biggl({1+z\over 1-z}G\biggr)\;\;. \eq
As usual \cite{akpa:3}, the poles of the function $G$ in $E$
are the only non-zero ${\cal A}(VV\to nH)$ amplitudes. It is easy to see
that ${\cal A}(VV\to nH)$ vanishes for $n\ge N+2$ provided that
\bq {4m_W^2\over m_H^2}=N(N+1)\;\;\; .\eq
Note that the nullification for longitudinal polarized bosons
starts from $n=N+2$ whereas in the case of transversely
polarized bosons this happens from $n=N+1$.
\par In the case where $V$ stands for $Z^0$ the analysis is exactly
the same and we arrive at the following condition:
\bq {4m_Z^2\over m_H^2}=N(N+1)\label{masz}\eq
where $N$ is again a positive integer.
\par The nullification of threshold amplitudes is the result
of very delicate cancellations between s$-$ and t$-$channel graphs.
In general, if such cancellations do not exist, one can take
the leading term in the Yukawa coupling which is the coherent
sum of the s$-$channel graphs. This leads to
violation of unitarity\cite{akpa:1}. The existence of specific relations
among the masses and/or the couplings could in principle restore the
high-$n$ unitarity. In order to see what the effect of cancellations
beyond threshold is, we calculate the amplitudes for the
processes $H(p_1)+H(p_2)\to H(p_3)+nH(k)$ in a $\phi^m$
scalar theory. The recursion formula is given by (see \fig{fi03}
for $\phi^4$)
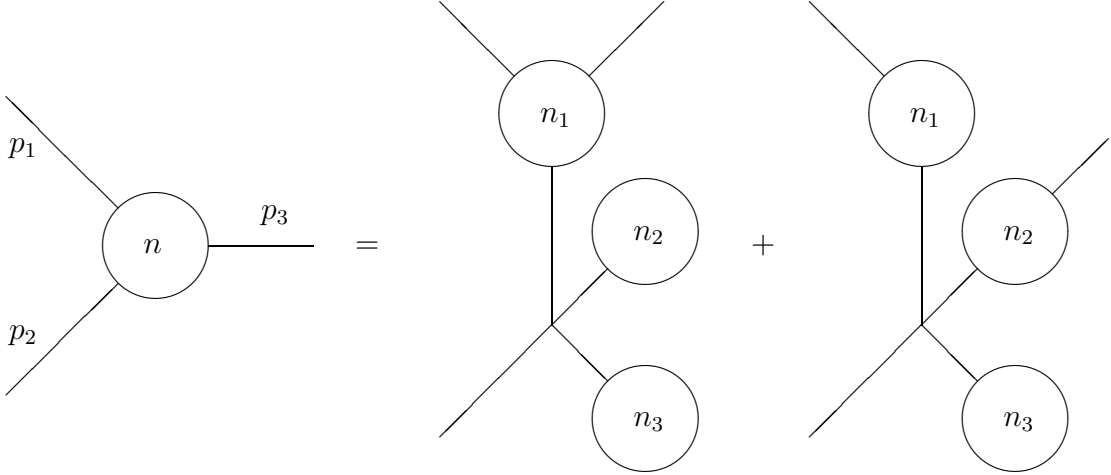
\begin{figure}[th]
\begin{picture}(16000,22000)

\drawline\fermion[\NE\REG](0,5000)[6000]
\global\advance\pmidx by -2000
\put(\pmidx,\pmidy){$p_2$}

\global\advance\pbackx by 1414
\global\advance\pbacky by 1414
\put(\pbackx,\pbacky){\circle{4000}}
\global\advance\pbacky by -300
\global\advance\pbackx by -450
\put(\pbackx,\pbacky){$n$}
\global\advance\pbackx by 450
\global\advance\pbacky by 300

\global\advance\pbackx by -1414
\global\advance\pbacky by 1414
\drawline\fermion[\NW\REG](\pbackx,\pbacky)[6000]
\global\advance\pmidx by -2000
\put(\pmidx,\pmidy){$p_1$}
\global\advance\pfrontx by 1414
\global\advance\pfronty by -1414

\global\advance\pfrontx by 2000
\drawline\fermion[\E\REG](\pfrontx,\pfronty)[4000]
\global\advance\pmidy by 1000
\put(\pmidx,\pmidy){$p_3$}

\global\advance\pfrontx by 6000
\global\advance\pfronty by -300
\global\advance\pfrontx by -450
\put(\pfrontx,\pfronty){$=$}
\global\advance\pfrontx by 450
\global\advance\pfronty by 300

\global\advance\pfrontx by 7000
\global\newcount\px
\global\newcount\py
\px=\pfrontx
\py=\pfronty

\drawline\fermion[\N\REG](\pfrontx,\pfronty)[3000]
\global\advance\pbacky by 2000
\put(\pbackx,\pbacky){\circle{4000}}
\global\advance\pbacky by -300
\global\advance\pbackx by -450
\put(\pbackx,\pbacky){$n_1$}
\global\advance\pbackx by 450
\global\advance\pbacky by 300

\global\advance\pbacky by 1414
\global\advance\pbackx by -1414
\drawline\fermion[\NW\REG](\pbackx,\pbacky)[4000]
\global\advance\pfrontx by 2828
\drawline\fermion[\NE\REG](\pfrontx,\pfronty)[4000]

\drawline\fermion[\S\REG](\px,\py)[3000]
\drawline\fermion[\SW\REG](\pbackx,\pbacky)[6000]

\drawline\fermion[\SE\REG](\pfrontx,\pfronty)[3000]
\global\advance\pbackx by 1414
\global\advance\pbacky by -1414
\put(\pbackx,\pbacky){\circle{4000}}
\global\advance\pbacky by -300
\global\advance\pbackx by -450
\put(\pbackx,\pbacky){$n_3$}
\global\advance\pbackx by 450
\global\advance\pbacky by 300

\drawline\fermion[\NE\REG](\pfrontx,\pfronty)[3000]
\global\advance\pbackx by 1414
\global\advance\pbacky by 1414
\put(\pbackx,\pbacky){\circle{4000}}
\global\advance\pbacky by -300
\global\advance\pbackx by -450
\put(\pbackx,\pbacky){$n_2$}
\global\advance\pbackx by 450
\global\advance\pbacky by 300

\global\advance\px by 8000
\global\advance\py by -300
\global\advance\px by -450
\put(\px,\py){$+$}
\global\advance\px by 450
\global\advance\py by 300
\global\advance\px by 6000

\drawline\fermion[\N\REG](\px,\py)[3000]
\global\advance\pbacky by 2000
\put(\pbackx,\pbacky){\circle{4000}}
\global\advance\pbacky by -300
\global\advance\pbackx by -450
\put(\pbackx,\pbacky){$n_1$}
\global\advance\pbackx by 450
\global\advance\pbacky by 300

\global\advance\pbacky by 1414
\global\advance\pbackx by -1414
\drawline\fermion[\NW\REG](\pbackx,\pbacky)[4000]

\drawline\fermion[\S\REG](\px,\py)[3000]
\drawline\fermion[\SW\REG](\pbackx,\pbacky)[6000]

\drawline\fermion[\SE\REG](\pfrontx,\pfronty)[3000]
\global\advance\pbackx by 1414
\global\advance\pbacky by -1414
\put(\pbackx,\pbacky){\circle{4000}}
\global\advance\pbacky by -300
\global\advance\pbackx by -450
\put(\pbackx,\pbacky){$n_3$}
\global\advance\pbackx by 450
\global\advance\pbacky by 300

\drawline\fermion[\NE\REG](\pfrontx,\pfronty)[3000]
\global\advance\pbackx by 1414
\global\advance\pbacky by 1414
\put(\pbackx,\pbacky){\circle{4000}}
\global\advance\pbacky by -300
\global\advance\pbackx by -450
\put(\pbackx,\pbacky){$n_2$}
\global\advance\pbackx by 450
\global\advance\pbacky by 300
\global\advance\pbackx by 1414
\global\advance\pbacky by 1414
\drawline\fermion[\NE\REG](\pbackx,\pbacky)[3000]

\end{picture}
\caption[.]{Diagrammatic representation of the recursion relation
for the process \lb $H(p_1)+H(p_2)\to H(p_3)+nH$.
The blob connected with one line corresponds to
the amplitude $H^*\to nH$ and that with two lines to the process
$H+H\to nH$.}
\label{fi03}
\end{figure}
\bqa
{a_3(n)\over n!}&=&
{-i\l\over q!}\sum
{ia_3(n_1)\over P_3(n_1) n_1!}
{ia(n_2)\over (n_2^2-1) n_2!}
\ldots
{ia(n_r)\over (n_r^2-1) n_r!}\nn\\
&+&
{-i\l\over (q-1)!}\sum
{ia_2(n_1;p)\over P_2(n_1;p_1) n_1!}
{ia_2(n_2;k)\over P_2(n_2;-p_3) n_2!}
\ldots
{ia(n_r)\over (n_r^2-1) n_r!}\label{recu}
\eqa
where $a_2(n;p)$ is the amplitude for the process
$H(p)+H(p')\to nH(k)$, $q=m-2$, $r=m-1$,
\bq P_2(n;p)=(p-nk)^2-1\;,\;\; P_3(n)=(p_1-p_3-nk)^2-1 \eq
and $k^2=p_i^2=1\;,\;i=1,2,3$.
\par Defining as usual
\bq f_3(x)=\sum_{n=Nq+q-1,N\ge 0} b_3(n)x^n \eq
where $a_3(n)=-in!P_3(n)b_3(n)$ and
\bq f_2(x;\w)=\sum_{n=Nq,n\ge 0}b_2(n;\w)x^n \eq
where $a_2(n;p)=-in!P_2(n;p)b_3(n;\w)$, with $\w=p\cdot k$,
we get
\bq
\biggl(\D^2-2\wab\D+\wab^2-\bw^2\biggr)f_3
={\l\over q!}f_3f_0^q+{\l\over (q-1)!}
f_2(\wa)f_2(-\wb)f_0^{q-1}\label{dife} \eq
with
\bq \wa=p_1\cdot k \;,\;\wb=p_3\cdot k\;,\; \wc=p_1\cdot p_3\;,\;
\wab=\wa-\wb\;,\;\bw=\sqrt{\wab^2-1+2\wc}
\eq
and
\bq f_2(\w)=(1+u^2)^{-s} F\biggl(-s,-s-{2\w\over q};1-{2\w\over q};-u^2
\biggr)\eq
with $u^2=-(\l/2m!)x^q$ and $u=e^\tau$ \cite{akpa:3}.
The solution of the \eqn{dife} is given by:
\bq f_3=-{u^{2\wab/ q}\over W}\biggl(
\ps{I} \int_{\tau}^{\infty} \ps{II}\F +
\ps{II}\int_{-\infty}^{\tau} \ps{I}\F
+C_1\ps{I}+C_2\ps{II}\biggr) \label{solu}
\eq
where the $C$'s are defined so as to make $f_3$ have an
expansion in integer powers of $x$, $\F$ is the inhomogeneous
part of \eqn{dife} multiplied by $4/q^2$,
and $\ps{I,II}$ are the two independent solutions of
\bq
\biggl( {d^2\over d\tau^2}-\w^2+{s(s+1)\over \cosh^2(\tau)}\biggr)\psi=0
\eq
with $s(s+1)={2m!/q^2 q!}$, $\w=2\bw/q$,
\bqa
\ps{I}&=&u^\w(1+u^2)^{-s}F(-s,-s+\w;1+\w;-u^2) \\
\ps{II}&=&u^{-\w}(1+u^2)^{-s}F(-s,-s-\w;1-\w;-u^2)
\eqa
and their Wronskian is $W=2\w$.
\par After some algebra (where we are interested only
in the terms in \eqn{solu} that have poles in $n-\bw-\wab$)
we get
\bq a_3(n)=-in! {2m!\over (q-1)!}\biggl( {\l\over 2m!}\biggr)^N
\sum_{k=0}^{N}\G_k {(a+N-k-1)!\over (N-k)!(a-1)!}\label{ampl}
 \eq
where
\bq \G_k={1\over k!}
\Biggl({d^k\over dx^k} \biggl( g(\bw;x)g(\wa;x)g(-\wb;x)\biggr)
\Biggr)_{x=0}\eq
\bq g(\w;x)\equiv F(-s,-s-{2\w\over q};1-{2\w\over q};x)\eq
$a=3s+2(q-1)/q$ and $n=Nq+q-1$.
For $s$ integer
the sum in \eqn{ampl} terminates, since $\G_k$ vanishes for
$k\ge 3s+1$. This termination
is the remnant of nullification beyond threshold. It shows that
the cancellations caused by the mass relations survive
beyond threshold. Of course an estimate of the real amplitude
(not only collinear configurations) is needed in order to
have a proof of unitarity restoration. Unfortunately this is
a rather difficult problem, since the definition of a lower bound
in analogy with the $H^*\to nH$ \cite{akpa:1}
amplitude, is no longer
possible because of the existence of destructive interference
between s$-$ and t$-$channel graphs. Nevertheless the persistence
of the cancellations beyond threshold
is a strong hint that unitarity might be
restored when nullification at threshold is present.
In any case the understanding of this phenomenon relies
on the recovery of the `symmetry' which is responsible for it,
which will guarantee the consistency of the theory.
\par Although a complete understanding of the nullification
phenomenon is  not
available, it is worth while to investigate
the consequences of Eqs.(\ref{masf},\ref{masw},\ref{masz})
for the masses of all elementary particles. The predictions
are summarized as follows:
\bq\begin{array}{cc}
\mathrm{Fermions} &\;\; m_f={N_f\over 2} m_H \\[12pt]
\mathrm{Bosons  } &\;\; m_V={\sqrt{N_V(N_V+1)}\over 2} m_H
\end{array}\eq
where $N_f$ and $N_V$ are integers. If we try to fit the
whole spectrum using these relations we need a super-light
Higgs, $m_H\le 2m_e$, where $m_e$ is the electron mass.
Besides the experimental exclusion of this possibility, such a
solution is not theoretically attractive. For instance one cannot
understand why only a few integers between 1 and $10^5$ ($N_f,N_W,N_Z$)
are realized in the spectrum.
On the other hand a solution of the form
\bq N_W=N_Z\;\;,\;\; N_f=0\;\;\mathrm{and}\;\; N_{top}\ge 1\eq
may be seen as a first-order approximation to the spectrum of
the existing elementary particles\footnote{For instance, we find,
for $m_Z\sim m_W=80$ GeV, that $m_H=67$ GeV ($N_{W,Z}=2$) and
$m_{top}=134$ GeV ($N_{top}=4$).}. In order to make the whole picture
phenomenologically relevant, one has to perform a more detailed
analysis \cite{akpa:n}. Recall that we
have relations between `bare' couplings that will be
changed by renormalization group equations, hopefully to a more
realistic form.
At any rate the existence of such relations could, in principle,
answer some important open problems of the Standard Model. For instance,
large mass splittings in the fermionic sector
(as in the case of the top quark),
can be explained in the context of \eqn{masf},
since this latter suggests
a kind of `quantization' of the fermion masses in terms
of the Higgs mass.
Of course, much more effort has to be spent in order to understand
these new relations. Nevertheless,
we can safely conclude that the high-$n$ limit
of multi-Higgs amplitudes is deeply related to the Higgs mechanism itself
and could
provide us with a new tool to understand the mass spectrum of the
elementary particles within the Standard Model.\\

\end{document}